\documentstyle[11pt,newpasp,epsf,twoside]{article}
\def\edcomment#1{\iffalse\marginpar{\raggedright\sl#1\/}\else\relax\fi}
\reversemarginpar
\begin{document}
\title{Cosmic Reionization and Galaxy Formation}
 \author{Masayuki Umemura, Taishi Nakamoto and Hajime Susa}
\affil{Center for Computational Physics, University of Tsukuba, Tsukuba, 
Ibaraki 305-8577, Japan}

\begin{abstract}
   Using 3D radiative transfer calculations on the reionization
of an inhomogeneous universe, QSO absorption line systems are simulated 
and they are compared with observations of Ly$\alpha$ continuum depression
at high redshifts.
By this comparison, it is found that the metagalactic UV
intensity decreases rapidly with $z$ at $z>4$ as 
$I_{21}=0.5\exp [3(4-z)]$, and the reionization must have taken place
between $z=6$ and 10. Based on this time-dependence of UV background
intensity, we explore the collapse of pregalactic clouds in 
the UV background, and find that the self-shielding is prominent 
above a mass scale as
$
M_{\rm BIF}=3.0\times 10^{11} M_\odot 
[(1+z_c)/5]^{-4.2}(I_{21}/0.5)^{0.6}.
$
This mass scale predicts the bifurcation of galactic morphology, 
and by confrontation with observations 
it turns out that the bifurcation mass successfully 
discriminates between elliptical and spiral galaxies. 
\end{abstract}

\section{Introduction}

   The cosmic reionization is an issue of great significance in cosmology
in relation to the formation of galaxies. The information 
on the ionization states has been accumulated by the observations 
of Ly$\alpha$ absorption lines in high redshift QSO or galaxy spectra. 
Recently, 3D cosmological hydrodynamic simulations 
(Cen et al. 1994; Miralda-Escude et al. 1996; Gnedin \& 
Ostriker 1996; Zhang et al. 1997) 
have revealed that the Ly$\alpha$ absorption 
systems can be explained in terms of the absorption by intergalactic density 
fluctuations. Also, the radiative transfer effects have been stressed
by 3D calculations on the reionization of intergalactic matter
(Abel, Norman, \& Madau 1999; Razoumov \& Scott 1999; Gnedin 2000; 
Nakamoto, Umemura, \& Susa 2000; Ciardi et al. 2000). 
Here, to elucidate the reionization history,
we simulate QSO absorption lines by using the results of 3D radiative transfer 
calculations and compare them to observations
at high redshifts. Also, the collapse of pregalactic clouds in a reionized
universe is explored in relation to the formation of protogalaxies
in UV background radiation.

\begin{figure}[t]
\begin{center}
\plotfiddle{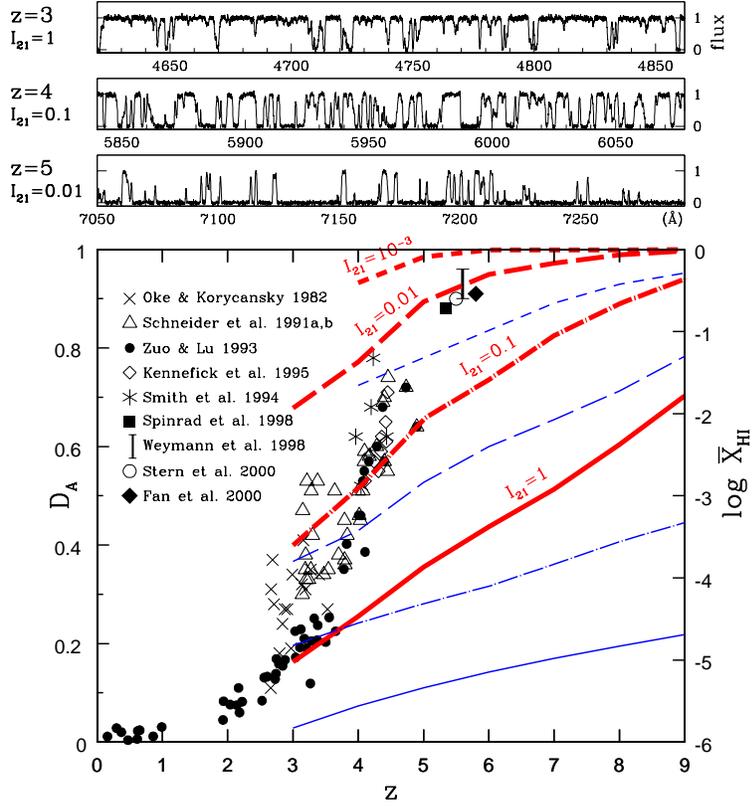}{10cm}{0}{55}{55}{-165}{-100}
\end{center}
\caption{The simulated Ly$\alpha$ absorption lines against wavelength 
at $z=3$ ({\it top panel}) with $I_{21}=1$, $z=4$ ({\it second}) 
with $I_{21}=0.1$, and $z=5$ ({\it third}) with $I_{21}=0.01$. 
The bottom panel is the diagram of 
Ly$\alpha$ continuum depression
({\it thick gray curves}) against redshifts. Symbols are
observations. 
Also, the mean neutral fractions $\overline{X}_{HI}$ ({\it thin curves}) are 
shown. The same line type corresponds to the same UV intensity.
}
\end{figure}

\section{QSO Absorption Lines at High Redshifts}

\subsection{Ly$\alpha$ Continuum Depression and Reionization History}

   In a simulation box, random Gaussian density fluctuations 
are generated based upon the Zel'dovich approximation
in the context of CDM cosmology.
The simulation box is irradiated by the isotropic UV 
background radiation of a power law-type spectrum, 
$I_\nu=I_{21}10^{-21}(\nu_L/\nu)$ 
erg cm$^{-2}$ s$^{-1}$ Hz$^{-1}$ sr$^{-1}$,
where $\nu_L $ is the Lyman limit frequency.
The ionization structure is obtained by solving the 
three-dimensional radiative transfer equation. 
The details of numerical techniques are presented in
Nakamoto, Umemura, \& Susa (2000) (see also the paper by Nakamoto
et al. in this volume).

   The resultant ionization degrees in the universe are different from place to 
place by more than three orders of magnitude. Due to such inhomogeneous 
ionization structure, relatively low ionization regions could produce strong 
absorption in quasar spectra. To make a direct comparison with the observations, 
we simulate absorption lines. 
First, we focus on Ly$\alpha$ absorption. 
To match the recent observations by the Keck telescope, 
we adopt the resolution of 
R=45000 and the variance of 0.04, and assume the Voigt profile of lines. 
The simulated Ly$\alpha$ absorption features are 
shown in upper three panels in Figure 1. 

In order to compare the simulated absorption degree with observations, 
we have assessed the so-called continuum depression, $D_A$.
In Figure 1, the simulated $D_A$ is compared to observations at high 
redshifts. We see that any model with a constant UV intensity does not match 
the observed trend that $D_A$ tends to grow quickly at redshifts
higher than 4.
This implies that the metagalactic UV intensity must decrease 
rapidly with $z$ at $z>4$ by two orders of magnitude at least. 
The redshift dependence of UV background is required to be
$I_{21}=0.5\exp [3(4-z)]$ at $z>4$.
If the UV intensity decreases in this fashion, 
the reionization epoch is estimated to be $z \approx 6$. 
If the intensity keeps a level of $I_{21} =0.01$ at $z>5$,
the reionization epoch is assessed
to be $z \approx 9$.  

Thus, it is concluded that the cosmic reionization must 
have taken place between $z=6$ and 10. However, the Ly$\alpha$ absorption is not 
appropriate to determine the reionization epoch more accurately, because, as seen 
in the absorption features in Figure 1, Ly$\alpha$ is too strongly 
depleted even if the 
mean neutral fraction is less than $10^{-2}$. 
In other words, $D_A$ is no longer sensitive to 
$X_{HI}$ around the cosmic reionization epoch.

\subsection{H$\alpha$ Forest}

   As shown in the previous subsection, Ly$\alpha$ has too high line 
opacity to probe the universe at $z>5$. Therefore, three conditions are 
required for a line in order to 
investigate the universe at $z>5$: 
(1) it has lower line opacity than Ly$\alpha$, 
(2) line emission is detectable, and 
(3) it has lower extinction against dust because young 
star-forming galaxies are often dust-enshrouded. The most favorable solution 
seems to be H$\alpha$ absorption lines. 
The H$\alpha$ absorption is likely to be relatively weak around $z=3$, 
while it could be sensitive to the ionization degrees at $z>4$. 
Therefore, 
H$\alpha$ forest can be a more powerful tool to probe the universe at $z>5$. 
H$\alpha$ forest has been never detected so far. The reason comes from the fact that 
H$\alpha$ has much weaker opacity than Ly$\alpha$. 
From observational points of view, the H$\alpha$ 
continuum depression can be detected by low-dispersion spectroscopy 
or narrow-band photometry. Furthermore, H$\alpha$ forest is subject 
to less UV bump effects for 
AGNs compared to Ly$\alpha$ forest. The wavelengths of H$\alpha$ forest drop on 
$3\mu {\rm m} \la \lambda_{{\rm H}\alpha} \la 7 \mu {\rm m}$ 
at $4\la z \la 10$. 
Thus, the observations can be done with Subaru IRCS, 
IRIS, SIRTF, NGST, or H2/L2. If one can obtain the absorption features with the 
resolution greater than 10000, one can recover the density fluctuations at high 
redshifts. They allow us to determine the linear amplitude of pregalactic 
perturbations which is by no means measured in the CBR due to the strong 
Sunyaev-Zeldovich effects in galactic scales. If one has the amplitude of linear 
density fluctuations at galactic scales, one can not only set 
the initial condition for 
galaxy formation, but also make more reliable determination of cosmological 
parameters.

\begin{figure}[t]
\begin{center}
\plotfiddle{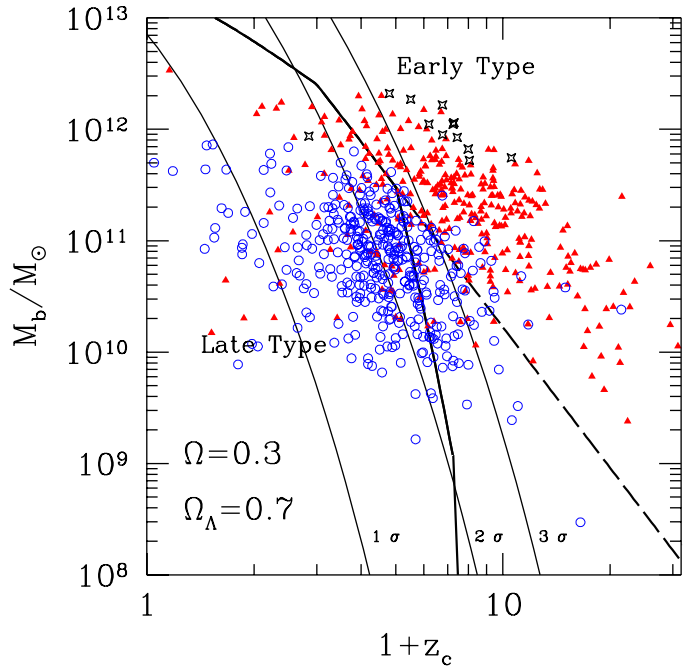}{10cm}{0}{50}{50}{-150}{-80}
\end{center}
\caption{Confrontation of the bifurcation theory to observations.
The thick solid and dashed lines represent
the bifurcation mass $M_{\rm BIF}$, respectively for
model A and B.
Small open circles represent
spiral galaxies. Small filled triangles
denote the E and E-S0 galaxies, and open stars are
massive elliptical galaxies from  X-ray observations.
Three thin solid lines marked as $1\sigma$, $2\sigma$, and $3\sigma$
are density fluctuations in the CDM cosmology. 
}
\end{figure}

\section{Galaxy Formation in UV Background Radiation}

\subsection{Bifurcation Theory}

Here, the cosmology is assumed to have $\Omega_0=0.3$,
$\Omega_\Lambda=0.7$, $h=0.7$, and $\Omega_{\rm b} h^2=0.02$
with usual meanings. 
Susa \& Umemura (2000a) have explored
the sheet collapse of a pregalactic cloud under UV background radiation,
with including H$_2$ chemistry.
They have found that the cloud evolution is specified 
by two initial parameters, i.e., 
the mean density ($\bar{n}_{\rm ini}$) and the thickness ($\lambda$).
The initial parameter dependence is translated into the baryonic mass 
[$M_{\rm b}\equiv (4\pi/3) \bar{n}_{\rm ini}(\lambda/2)^3$]
and the collapse epoch ($z_{\rm c}$) by 
assuming the initial stage is close to the maximum expansion
of a density fluctuation.
The numerical results have shown that the cloud evolution bifurcates
into two characteristic states, i.e.,
 (a) a pregalactic cloud is self-shielded
against the external UV in the course of the sheet collapse, so that
the cloud cools down below $10^3$K due to H$_2$ cooling
and undergoes efficient star formation.
Hence, it is expected to evolve into an early type galaxy with 
a high bulge-to-disk ratio (B/D) due to the dissipationless virialization,
or (b) a cloud is not self-shielded during the sheet 
collapse, but is self-shielded through the shrink to the rotation
barrier. This leads to the retarded star formation, and thus
the virialization would proceed in a fairly dissipative fashion. 
As a result, a late type
(small B/D) galaxy would be preferentially born. 
Such bifurcation is specified solely by a mass scale as
\begin{eqnarray}
M_{\rm BIF}=3.0\times 10^{11} M_\odot 
\left(\frac{1+z_c}{5}\right)^{-4.2}
\left(\frac{I_{21}}{0.5}\right)^{0.6} \label{eq:msb},
\end{eqnarray}
where $z_c$ is the collapse epoch and
$I_{21}$ is the UV background intensity in units of
$10^{-21}{\rm erg~ s^{-1} cm^{-2} str^{-1} Hz^{-1}}$.

\subsection{Confrontation with Observations}

Here we include the effect of the evolution of 
UV background radiation.
We assume $I_{21}=0.5\left[\left(1+z\right)/3 \right]^3$ for $z\le 2$ 
and $I_{21} =0.5$ for $2 < z \le 4$.
This dependence is consistent with the UV intensity
in the present epoch (Maloney 1993; Dove \& Shull 1994), and the value inferred from 
the QSO proximity effects at intermediate redshifts 
(Bajtlik, Duncan, \& Ostriker 1988; Giallongo et al. 1996).
As for $z > 4$, two models are employed. 
The first one is (A) the exponentially damping model,
$I_{21}=0.5\exp\left[3\left(4-z\right)\right]$, which is inferred by
the continuum depression, and 
the second one is (B) the constant extrapolation model,
$I_{21}=0.5$. 

Then, based upon the above paradigm of the galaxy formation under UV background,
the evolutionary bifurcation of pregalactic clouds is 
confronted with observations of elliptical and spiral galaxies.
Using the observed properties of galaxies, the collapse epochs are 
assessed for each type of galaxies with attentive mass estimation
(Susa \& Umemura 2000b). 
The comparison of observed galaxies with the bifurcation theory 
is presented in Figure 2.
Small open circles represent
spiral galaxies in Persic \& Salucci (1995).
Small filled triangles
denote the E and E-S0 galaxies in Faber et al. (1989).
Open stars are
elliptical galaxies from  X-ray observations 
in Matsumoto et al. (1997).
The thick solid and dashed lines represent
the bifurcation mass $M_{\rm BIF}$, respectively for
model A and B, where both are identical at $z_{\rm c}<4 $.
By this direct comparison of the theory with the observations, 
it turns out that  
the theoretical bifurcation branch successfully discriminates
between elliptical and spiral galaxies.
This suggests that the UV background radiation 
could play a profound role for the differentiation
of the galactic morphology into the Hubble sequence.
In Figure 2, the density fluctuations in the CDM cosmology
are also plotted  by three thin solid lines marked as 
$1\sigma$, $2\sigma$, and $3\sigma$, where $\sigma$ is
the variance of CDM perturbations. We find that
the bifurcation mass scale coincides with the $(2-3)\sigma$
density fluctuations. Thus,
the present results are also intriguing from a view point of
the density morphology relation of galaxies, because
higher $\sigma$ peaks reside preferentially in denser regions 
rather than in low-dense regions. 

\section{Conclusions}

   Using the ionization structure in an inhomogeneous universe 
obtained by 3D radiative transfer calculations on the reionization,
we have simulated QSO absorption lines at high redshifts.
By comparison of simulated Ly$\alpha$ continuum depression with 
recent observations, it has been found that the metagalactic UV
intensity must decrease rapidly with $z$ at $z>4$, 
and the reionization must have taken place
between $z=6$ and 10. Based on this time-dependence of UV background
intensity, we have explored the collapse of pregalactic clouds in 
the UV background, and found that the degree of self-shielding bifurcates
the evolution of protogalaxies. The bifurcation is characterized
by a mass scale dependent on UV background intensity and redshift.
By confrontation with observations 
it has turned out that the bifurcation mass successfully 
discriminates between elliptical and spiral galaxies. 
This suggests that the UV background radiation 
is closely related to the final bulge-to-disk ratios
of galaxies.

We have assumed here AGN-like ionizing sources, but 
which type of sources reionized the universe is
still an issue under a lot of debate. Anyhow, the present analysis
suggests that to elucidate precisely the reionization history 
is extraordinarily important for the understanding of galaxy
formation.

\acknowledgments
The numerical analysis has been made with computational facilities 
at Center for Computational Physics in University of Tsukuba.

\end{document}